\documentclass[aps,preprint,amsmath]{revtex4-1}

\usepackage{hyperref}

\usepackage [english]{babel}
\usepackage{graphicx}
\usepackage{dcolumn}
\usepackage{bm}
\usepackage{latexsym}
\usepackage{mathtools}
\usepackage{array}

\begin{document}

\title{Mode switching in ring lasers with delayed optical feedback.}

\author{Dontsov A.A.}

\begin{abstract}
We have demonstrated that a rather weak external optical feedback with delay can lead to the mode switching of the counterpropogating modes.  The delay time should be longer then any system characteristic time. The equations describing the ring resonator with the delayed optical feedback was obtained from the first principles.  The analytical formula for the necessary external feedback strength was derived. Other than the regular behavior, the system demonstrates a complex chaotic behavior.
\end{abstract}

\affiliation{Ioffe Physical-Technical Institute of the Russian Academy of Sciences,26 Polytekhnicheskaya, St. Petersburg 194021, Russian Federation}
\email[E-mail me at: ]{operatorne@yandex.ru}

\maketitle

\section{Introduction.}
The ring lasers are actively studied recently\cite{ActSt1,ActSt2,ActSt3,ActSt4,ActSt5}.
The common counterpropogating mode switching effect(without feedback) and other interaction effects in the ring lasers was predicted theoretically for the two-level active media \cite{Zeghlache}. In some explanations of this effect the assumption about the substantial effect of noise was used \cite{Noise1,Noise2}. The mode switching effect have been observed experimentally in the gas lasers \cite{Tang,Gas2}, in the dye lasers \cite{Dye} and in the Nd-YAG laser \cite{Nd-YAG}. Frequency detuning between the center of active-media gain spectrum and the resonator-mode frequency is a very important parameter for the mode switching effect in the two-level active media. For large enough frequency detuning the unidirectional regime becames unstable and superseded by the mode switching regime \cite{Zeghlache}.

\begin{figure}[]
\center{\includegraphics[width=1\linewidth]{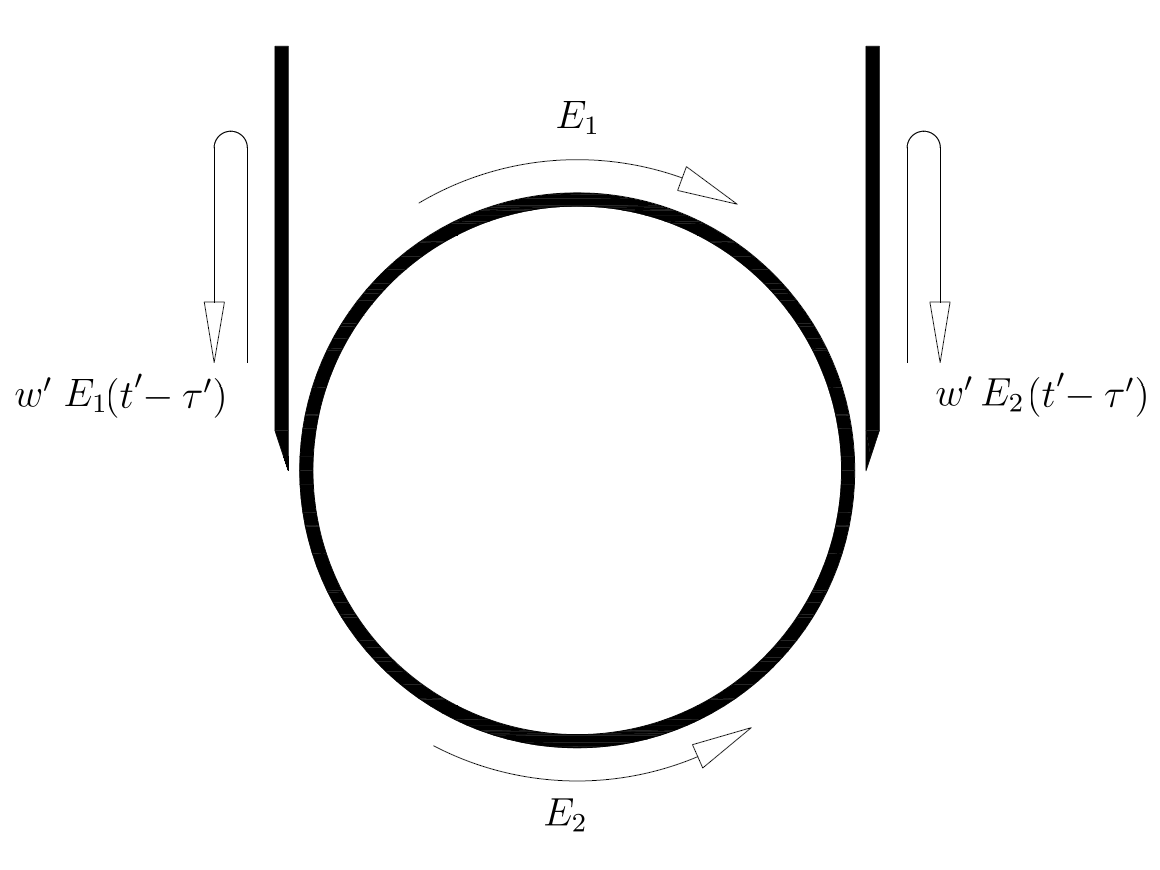}}
\caption{The ring laser with the delayed optical feedback.}
\label{fig:scheme}
\end{figure}

 The purpose of this paper is to show that it is possible to achieve the mode switching even when the frequency detuning is too small to induce it. This can be achieved by adding an external optical feedback as illustrated in fig. \ref{fig:scheme}. We consider just two counterpropagating modes: "plus-" and "minus-mode" with the same frequency $\omega_c$. The part of "plus-mode" ("minus-mode") power leaves the resonator, propagating outside the cavity during the time $\tau'$ and than return back to the opposite mode - "minus-mode" ("plus-mode").

 The first consideration of the semiconductor ring laser with the delayed optical feedback was done in \cite{conc}. The model equation of such a laser was written using a common semiconductor ring-laser equations \cite{sorel}.

 The second space harmonic of the population inversion was not considered because of a large carriers diffusion. As a result, some interesting dynamical effects (even the chaotic behavior) was obtained, except for the mode switching due to the optical feedback \cite{conc}. In this paper the equation of the ring laser with delayed optical feedback will be obtained from the first principles taking in account the second space harmonica. As it will be shown below, this equations coincide with that from \cite{conc} for the high carrier diffusion. When the diffusion is small as it is in the case of gas laser or quantum dot semiconductor laser, the mode switching effect could be achieved by the adding of the external delayed optical feedback.

\section{The equations for the ring resonator with the external delayed optical feedback.}
The device with external delayed optical feedback does not described by the common semi-classical laser equations \cite{Haken} and first of all we should derive the proper equations. The laser semi-classical equation ( with notation from \cite{nota} ) can be written as

\begin{subequations}
\label{1eq:poln}
\begin{eqnarray}
\label{equationa}
i(\partial_{t'}+\mathcal{K}_j) \beta_j(t')-\omega_{c\,j} \beta_j(t')=\nonumber\\\frac{n}{L}\int_0^L dx g_j^*(x)\alpha(x,t')
\end{eqnarray}
\begin{eqnarray}
\label{equationb}
i(\partial_{t'}+\gamma_\bot) \alpha(x,t')-\omega_A \alpha(x,t')=\nonumber\\-\mathfrak{D} (x,t') \sum_j g_j(x) \beta_j(t')
\end{eqnarray}
\begin{eqnarray}
\label{equationc}
i(\partial_{t'}+\gamma_\|) \mathfrak{D}(x,t')=\nonumber\\i \sigma \gamma_\|+2 \sum_j[g_j(x)\alpha^*(x,t') \beta_j(t')-c.c.)]
\end{eqnarray}
\end{subequations}


Here the electric field expanded in the modes of resonators $E_{in}(x,t')= \sum \beta_j(t') \, g_j(x)\,c_j$, where index j takes values of all mode numbers.

$\beta_j(t')$ is a mode complex amplitude and $g_j(x)$ is a space distribution of the mode field,
$c_j$ are the normalization constants.

The value $\alpha(x,t')$ is the induced polarization, $\mathfrak{D}(x,t')$ is the population inversion. $\mathcal{K}_j$ is the the resonator loss, $\gamma_\bot$ and $\gamma_\|$ are the cross- and longitudinal- relaxation times, $n$ is the active-medium particle concentration, $\omega_{c\,j}$ is the frequency of j resonator`s mode and $\omega_A$ is the central frequency of active-media gain band. The symbol  $\sigma$ means the steady value of $\mathfrak{D}(x,t')$ for the zero fields ($\beta_j(t')=0$; $\partial_{t'} \mathfrak{D} (x,t')=0$).

To explain the derivation of the equations with the delay optical feedback, let us consider the simplest case: a single mode system without am active medium. Let us suppose that there are external independent fields exciting the resonator mode. This fields $E_{ex}(x,t')=f_{ex}(t') \widehat E_{ex}(x) $ satisfy  the wave equation in free space $\Delta E_{ex}(x,t')-\frac{1}{c^2} \frac{\partial^2 E_{ex}(x,t')}{{ \partial t'}^2}=0$. The resonator is a "perturbation$"$ of the free space. The spatial part $\widehat E_{m}(x)$ of a single mode field $E_{m}(x,t')=f_{m}(t') \widehat E_{m}(x)$ in the resonator satisfies the equation $\Delta \widehat E_{m}(x)+\epsilon(x)\frac{ \omega_m^2}{c^2} \widehat E_{m}(x)=0$. In an open resonator $\omega_m$ should be a complex number, but it is not significant. Our purpose is to obtain the solution of the total equation  $\Delta E_{total}(x,t')-\frac{\epsilon(x)}{c^2} \frac{ \partial^2 E_{total}(x,t')}{{\partial t'}^2}=0$ with the "boundary condition$"$ that the field far from the resonator is $E_{total}(\infty,t')=E_{ex}(\infty,t')$. So, let us look for the total solution in the approximate form $E_{total}(x,t')=f_{m}(t') \widehat E_{m}(x)+f_{ex}(t') \widehat E_{ex}(x) $. Here  $f_{m}(t')$ is the only unknown function. Using all conditions mentioned above and the $E_{m}(x)$ normalization $\int \widehat E_{m}(x) \epsilon(x) \widehat E^*_{m}(x)=1$, one obtains $\omega_m^2  f_m(t')+\ddot f_m(t')+  \ddot f_{ex}(t') \alpha=0$ ,where $\alpha=\int \widehat E_{ex}(x)(\epsilon(x)-1) \widehat E^*_{m}(x)$. This expression is equivalent to the linear oscillator equation with an external force.

 The next step is to exclude the second time derivation. We suppose the external force to be almost in the resonance $\omega_m \sim \omega_{ex}$.  If the external field has the form $f_{ex}(t')=F(t') E^{-i \omega_{ex} t'}$, where $F(t')$ is a slowly varying function, we can write $f_m(t')= C(t') E^{-i \omega_{m} t'}$. Here $C(t')$ is a slowly varying function. Using the slow variation of functions and assuming $\ddot C(t') \rightarrow 0$ , $\ddot F(t') \rightarrow 0$ and  $\alpha \dot F(t') \rightarrow 0$ one obtains  $\dot f_m(t')=-i \omega_m f_m(t')+i\frac{\alpha\omega_{ex}^2}{2 \omega_m} f_{ex} (t')$. So, the external force leads to the simple summand in the right part.

The similar calculations for the two coupled modes with the active medium \cite{Haken} leads to the modified form of (\ref{1eq:poln}). After some mathematical transformation that are similar to that in \cite{Zeghlache} one obtains:

\begin{equation}
\label{eq:osn_zader}
(\partial_{t}+1) E_1=w E_{2 \tau}+(1+i \Delta) \tilde{A}(E_1 D_0+E_2 D_1^*)
\end{equation}

$$(\partial_{t}+1) E_2=w E_{1 \tau}+(1+i \Delta) \tilde{A}(E_2 D_0+E_1 D_1)$$
$$(\partial_{t}+d_\|)D_0=d_\|-\tilde{d_\|} D_0(|E_1|^2+|E_2|^2)-\tilde{d_\|}(E_1 E_2^* D_1+cc)$$

$$(\partial_{t}+d^a_\|)D_1=-\tilde{d_\|} D_1(|E_1|^2+|E_2|^2)-\tilde{d_\|}E_1^*E_2 D_0 .$$

Here
w - is a small coupling parameter,
$E_j(t)=(\frac{4 {|g|}^2}{\gamma_{\|} \gamma_{\bot}})^{1/2} e^{i \omega_c t} \beta_j(t)$;
$\quad D_n(t)=\frac{1}{L \sigma} \int_0 ^L \mathfrak{D} e^{i 2 n k x}dx$;
$\quad t=t' \mathcal{K}_1$;
$\quad \tau=\tau' \mathcal{K}_1$;
$\quad A=\frac{\sigma n |g|^2}{\gamma_\bot \mathcal{K}_1}$;
$\quad d_\|=\frac{\gamma_\|}{\mathcal{K}_1}$;
$\quad d_\bot=\frac{\gamma_\bot}{\mathcal{K}_1}$;
$\quad  w =w' e^{i \omega_c \tau_d }$;
$\quad |g|^2=\frac{2 \pi \omega_c}{\hbar}\,\mu^2$; $\Delta=\frac{\omega_{c}-\omega_A}{\gamma_\bot}$ ; $\mu$ is a dipole moment. Subscrip "$\tau$" means delaying on time $\tau$, for example $E_{1 \tau}=E_{1}(t- \tau)$. Symbol "$\sim$" over the other symbol means $\tilde f=\frac{f}{1+\Delta^2}$. Symbol "cc" is a short-cut for complex conjugation. Generally, the constant  $d^a_\|\neq d_\|$, it allows to takes into account the additional relaxation of inversion grating due to carriers diffusion \cite{Diffus}.

 Note that the energy that returns back to the resonator over the feedback was taken from one of the resonator`s modes. This losses should be taken into account in the $\mathcal{K}_j$ constants. The equations (\ref{eq:osn_zader}) describes the main types of ring lasers with the delayed optical feedback. It is the main equations of the paper and all the following are based on them.

If one neglects the delayed feedback ($w=0$), the equations (\ref{eq:osn_zader}) reduce to that from \cite{Zeghlache}. On the other hand, for the common semiconductor laser the diffusion coefficient is large $d^a_\|>>d_\|$  and the spatial harmonica $D_1(t)$ could be adiabatically excluded. Assuming $D_1'(t)=0$ we obtain $d^a_\| D_1\approx-\tilde{d_\|}E_1^*E_2 D_0$  and the equations (\ref{eq:osn_zader}) became similar to the equations from  \cite{conc}. However, there is a difference between them in the terms like $\sim |E_1|^2 E_1$ and $\sim |E_1|^2 |E_1|^2$. This difference is due to the fact that we neglected the possible dependence of constants in (\ref{1eq:poln}) from the light intensity.We also neglected the non delayed backscattering. Both effects are not very significant for the mode switching effect.
\section{The numerical study of the mode switching effect.}
\begin{figure}
\begin{minipage}[h]{0.45\linewidth}
\center{\includegraphics[width=0.9\linewidth]{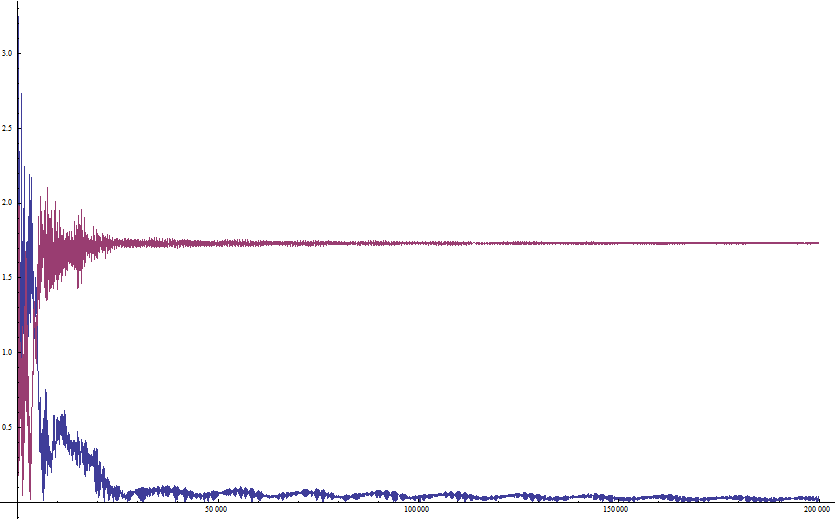} \\ a)}
\end{minipage}
\hfill
\begin{minipage}[h]{0.45\linewidth}
\center{\includegraphics[width=0.9\linewidth]{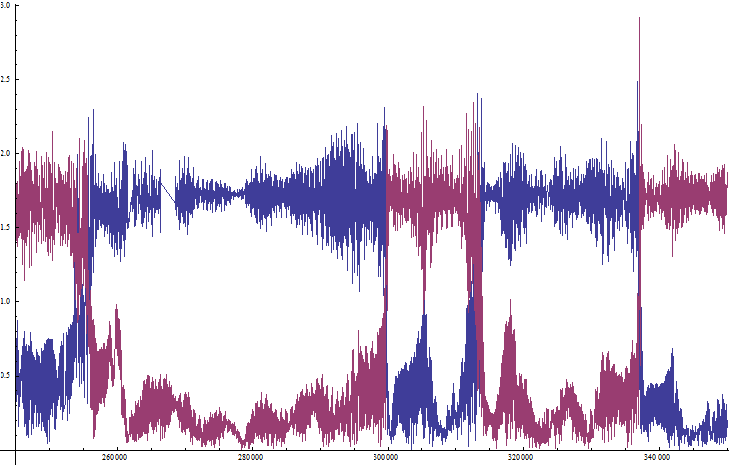} \\ b)}
\end{minipage}
\begin{minipage}[h]{0.45\linewidth}
\center{\includegraphics[width=0.9\linewidth]{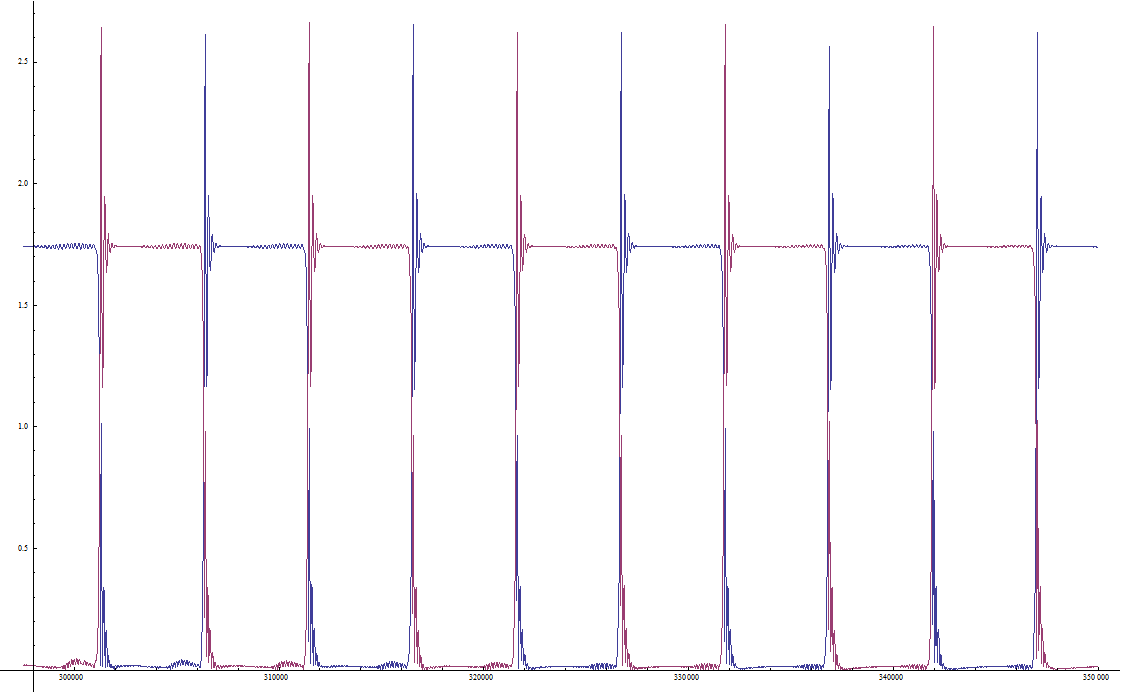} \\ c)}
\end{minipage}
\caption{Numerical solution of the equations (\ref{eq:osn_zader}) for the various feedback strength $w$. (a) $w$=0.0025 (b) $w$=0.0039 (c) $w$=0.01}
\label{fig:Evol}
\end{figure}

 The numerical solutions of (\ref{eq:osn_zader}) for the $CO_2$ laser with typical parameters $A=4$ , $d_\|=d_\|^a =2\cdot 10^{-4}$, $\Delta=0.02$, $\tau=5000$ and different $w$ are represented in the fig. \ref{fig:Evol}. According to \cite{Zeghlache} $ \Delta_{crit}\approx\sqrt{\frac{d_\| A^2}{A-1}}=0.032$, so there is no mode switching without external feedback($w=0$)

It is shown in the pictures, for a small feedback strength $w=0.0025$ the unidirectional steady state solution is stable. For the critical value of the feedback strength $w_{crit}=0.0039$ this solution loses stability and a complex chaotic behavior is observed.There is the stable mode switching regime in the system when the feedback strength is high enough ($w=0.01$).

The pattern  "unidirectional solution - chaotic behavior - stable mode switching$"$ is the typical pattern for the growing w and the different parameter values. The chaotic behavior is observed in a narrow range of $w$. However, the behavior of the system is very complex and we don't claim for it total investigation.

 \section{The analytical formula for the threshold feedback strength $w$.}

Now our purpose is to find the formula for the feedback strength $w_{switch}$ that force the laser to operate in a mode switching regime. As it can be seen from above, $w_{switch}$ has the numerical value close to $w_{crit}$ wherein the unidirectional solution loses it stability $w_{switch}\sim w_{crit}$.

To find  $w_{crit}$ we shall investigate the stability of the unidirectional solution of the equations (\ref{eq:osn_zader}). These equations are not the common nonlinear differential equations but the differential equations with a deviating argument. This type of equations is scantily studied , but the solution-stability problem was discussed in great details (see for example \cite{Eltsgol}).

The calculation of $w_{crit}$ is represented in Appendix \ref{AppWcalc} and the approximate upper bound for the $w_{crit}$ is
 \begin{equation}
\label{eq:result}
|w_{crit}|^2 = \frac{\sqrt{A-1}\,\sqrt{d_\|}([(A-1)d_\|+d_\|^a]^2-\Delta^2 (A-1)\,d_\|)}{d_\|^a+(A-1)d_\|}
\end{equation}
for $d_\|<<1$,  $d^a_\|<<1$, $\Delta^2<<1$ and $\tau>>1/d_\|$.

When $d_\|=d_\|^a$ the expression is simplified:
 \begin{equation}
\label{eq:resultUPRO}
|w_{crit \,simp}|^2 = \frac{\sqrt{A-1}\,\sqrt{d_\|}(A^2\,d_\|-\Delta^2(A-1))}{A}.
\end{equation}

\begin{figure}[]
\center{\includegraphics[width=1\linewidth]{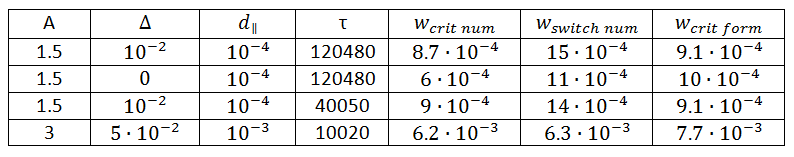}}
\caption{Comparation of the formula (\ref{eq:resultUPRO}) and the numerical calculations results.The values $w_{crit\,\,num}$ and $w_{switch\,\,num}$ are the critical feedback force and the feedback strength that force the laser to operate in a stable switching regime, obtained due to the numerical calculations. The $w_{crit\,\,form}$ is the critical strength calculated, using the formula (\ref{eq:result}).}
\label{fig:table}
\end{figure}

When $w>w_{crit}$  guaranteed that the steady unidirectional solution lose it`s stability. The accuracy of (\ref{eq:result}) was checked by comparison of the values $w_{crit}$ and "stable switching" strength $w_{switch}$ obtained by the accurate numerical solution of (\ref{eq:osn_zader}) with the corresponding value $w_{crit}$ obtained from (\ref{eq:result}). The results are represented in fig. \ref{fig:table}

The interesting feature of this formula is that when $\Delta=0$, $w_{crit} \neq0$. It means that even when there is no frequency detuning, the mode switching regime is possible. In physical parameters described after (\ref{1eq:poln}) this inequality is

\begin{equation}
w^2>\frac{1}{\gamma_\bot}\sqrt{\frac{\sigma n |g|^2}{\gamma_\bot \mathcal{K}_1}-1}\sqrt{\frac{\gamma_\|}{\mathcal{K}_1}}\frac{\sigma n |g|^2 \gamma_\|}{\mathcal{K}_1^2}
\end{equation}
 for $\Delta=0$. For the semiconductor laser with quantum dots (without carrier diffusion and detuning) $d_\|=3\cdot 10^{-3}$;
$\quad d_\bot=3$ ;
$\quad A=2$ ;
$\quad \Delta=0$, \cite{quantDots}  and the formula (\ref{eq:result}) gives $w>0.07$. Thus, only 7$\%$ of the leaking energy have to be coupled to the opposite mode. This gives a hope that the effect could be observers in such lasers.

\section{Conclusion}
In this paper the equations describing the ring resonator with the delayed external optical feedback (\ref{eq:osn_zader}) was obtained. The numerical study of these equations shows that a weak external delayed optical feedback is enough to induce the mode switching effect. The analytical formula for the critical value of the feedback strength (\ref{eq:result}) was obtained and the comparation with the numerical solutions shows the correctness of the formula (\ref{eq:result}). This formula could be applied to the semiconductor ring lasers with quantum dots, if one put $\Delta=0$. The calculation shows that it is theoretically possible to obtain mode switching effect in such lasers.

The equations of the model (\ref{eq:osn_zader}) have a very complicated behavior and reveal the different non trivial effects, but we do not claim for the full classification of these effects. For instance the solution could depend on the initial conditions. So, there are some different attractors in the system and each of them forms a basin of attraction. Some kind of "period doubling" effect was observed when the feedback force becomes strong $w\sim10 w_{crit}$, etcetera.

\begin{acknowledgments}
The author would like to thank A.M. Monakhov for the useful discussions.
\end{acknowledgments}

\appendix
\section{Derivation of the analytical formula for the feedback strength $w_{crit}$ enough to unidirectional steady state solution lose it`s stability.}
\label{AppWcalc}

First of all, let us rewrite the eqations (\ref{eq:osn_zader}) in real numbers. Making the replacment
$\quad E_{1,2}(t)=|E_{1,2}(t)| e^{i \varphi_{1,2}(t)}$;
$\quad D_1=|D_1|e^{i \varphi_s}$;
$\quad D_0=|D_0|$,
one obtains:
\begin{widetext}
\begin{eqnarray}
\label{eq:osn_zader_veshestv}
\dot{|E_1|}=-|E_1|+w |E_{2 \tau}| cos(\varphi_{2 \tau}-\varphi_{1})+\widetilde{A}|E_1| |D_0|+\gamma_m \widetilde{A}|E_2|\,|D_1| cos(\varphi_\Delta+\gamma_a)
\end{eqnarray}
\begin{eqnarray}
\nonumber
\dot{\varphi_1}|E_1|=w |E_{2 \tau}| sin(\varphi_{2 \tau}-\varphi_{1})+\Delta \widetilde{A} |E_1| |D_0|+\gamma_m\,\widetilde{A} |E_2| |D_1| sin(\varphi_\Delta+\gamma_a)
\end{eqnarray}
\begin{eqnarray}
\label{eq:osn_zader_veshestv}
\dot{|E_2|}=-|E_2|+w |E_{1 \tau}| cos(\varphi_{1 \tau}-\varphi_{2})+\widetilde{A}|E_2| |D_0|+\gamma_m \widetilde{A}|E_1|\,|D_1| cos(-\varphi_\Delta+\gamma_a)
\end{eqnarray}
\begin{eqnarray}
\nonumber
\dot{\varphi_2}|E_2|=w |E_{1 \tau}| sin(\varphi_{1 \tau}-\varphi_{2})+\Delta \widetilde{A} |E_2| D_0+\gamma_m\,\widetilde{A} |E_1| |D_1| sin(-\varphi_\Delta+\gamma_a)
\end{eqnarray}
\begin{eqnarray}
\nonumber
\dot{\varphi_s} D_0=-d_{\|} D_0 +d_{\|}-\tilde d_{\|} D_0 (|E_1|^2+|E_2|^2)-\tilde d_{\|} |D_1| |E_1| |E_2| 2 cos(\varphi_\Delta)
\end{eqnarray}
\begin{eqnarray}
\nonumber
\dot{D_1}=-d^a_{\|} D_1-\tilde{d_\|} D_1 (E_1^2+E_2^2)-\tilde{d_\|} D_0 E_1 E_2 cos(\varphi_\Delta)
\end{eqnarray}
\begin{eqnarray}
\nonumber
\dot{\varphi_s} D_1=-\tilde d_{\|} D_0 E_1 E_{2 \tau} sin(\varphi_\Delta)
\end{eqnarray}
\end{widetext}
Here $\varphi_\Delta=\varphi_2-\varphi_1-\varphi_s$;\quad $\gamma_m=\sqrt{1+\Delta^2}$ and $\gamma_a=Atan(\Delta)$. Generally, the constant $w$ is a complex number, but here we assume it real. For our purposes it is enough. The modules could be steady only if the phases satisfy the equality $\varphi_{1,2}=\Omega t+\psi_{1,2}$, where $\psi_{1,2}$ are constants, $\Omega \approx \Delta$ is an unknown constant, and the phase $\varphi_{s}(t)=\psi_s$ is a constant too.
The unidirectional steady state in the linear approximation in $w$ is

 $\quad D^{s}_0=\frac{1}{\tilde{A}}$;
 $\quad D^{s}_1=\frac{w}{\tilde{A} \gamma_m}$;
 $\quad E^{s}_1=\sqrt{A^2-\gamma_m^2}$;
 $\quad E^{s}_2=w\,\frac{d_\|^{a}+d_\| {E^s_1}^2}{\tilde d_\| E^s_1 \gamma_m}$;
 $\quad \psi_{2}-\psi_{1}-\psi_{s}=\pi$;
 $\quad \psi_{2}-\psi_{1}+\Omega \tau+\gamma_a=0$;

  The steady-state phase equalities have a simple physical meaning. The relation $\psi_{2}-\psi_{1}-\psi_{s}=\pi$ is equivalent to $\varphi_{2}(t)- \varphi_{2}(t)- \varphi_{s}(t)=\pi$. It means that the standing wave formed by counterpropogating modes, burns the population inversion grating, so the second space harmonica $D_1$ is shifted in phase on $\pi$ relatively to the standing wave.

  The relation  $\psi_{2}-\psi_{1}+\Omega \tau+\gamma_a=0$ is equivalent to $\varphi_{2}(t)- \varphi_{1}(t-\tau)=-\gamma_a$. It means that the phase of $E_2$ is almost equal to that for $E_1(t-\tau)$.

  To find the stationary value of $\Omega$ approximately is incorrect when $\Omega \tau >> 1$. It is a problem, because $w_{crit}$ is a rapidly oscillating function of the $ \tau$. So, to obtain the upper bound of $w_{crit}$, one can slightly change the $\tau$ to the nearest $\tau_{up}$ corresponding to the local maximum of $w_{crit}(\tau)$

To suppress the mode switching, the standing wave formed by the delayed counterpropogating modes should be in the phase with the grating burned by the non delayed standing wave. In other words $\varphi_{1}(t-\tau_{up})-\varphi_{2}(t-\tau_{up})-\varphi_{s}(t)=0$, i.e. the phase shift is zero. Using this condition and another steady state conditions, one obtains $\Omega\,\tau_{up}=\frac{\pi}{2}+\pi\,m$, where $m=...,-1,0,1,...$.


 Now let us consider the stability of unidirectional solution. The linearization of (\ref{eq:osn_zader_veshestv}) near the stationary point gives:
 \begin{widetext}
\begin{eqnarray}
\nonumber
\label{lineariz}
\dot{\delta E_1}=
-B_1 \cdot \delta E_2+
w \,cos(\psi_D-\Omega \tau) \cdot \delta E_{2 \tau} -\frac{B_4 B_1}{B_2} \cdot \delta D_1+
B_4 E_1^s \cdot \delta D_0
\end{eqnarray}
\begin{eqnarray}
\nonumber
\dot{\delta E_2}=
-B_1 \cdot \delta E_1+
B_1 \cdot \delta E_{1\tau}+
B_1  E_1^s \Delta \cdot \delta \psi_1-
B_1 E_1^s \Delta \cdot \delta \psi_{1\tau}-
B_4  E_1^s \cdot \delta D_1+\\ \nonumber
B_1 E_1^s \Delta \cdot \delta \psi_s +
(B_1 B_4 /B_2) \cdot \delta D_0
\end{eqnarray}
\begin{eqnarray}
\nonumber
\dot{\psi_2}=
B_4 \Delta \cdot \delta D_0 -
 B_2  \Delta \cdot \delta E1+
 B_2  \Delta \cdot \delta E_{1\tau}-
 \frac{(B_2 B_4  \Delta E_1^s)} {B_1} \cdot \delta D1 -
  B_2 E_1^s \cdot \delta \psi_1 +
   B_2 E_1^s \cdot \delta \psi_{1 \tau} -\\
    B_2  E_1^s \cdot \delta \psi_s+
     (2 B_2^2  E_1^s cos[1/2 (\gamma_a+ \psi_s + \Omega \tau)]/B_1) \cdot \delta E_2
\end{eqnarray}
\begin{eqnarray}
\nonumber
\dot{\psi_1}=
B_4 \Delta \cdot \delta D_0 -
(B_1 B_4  \Delta)/(B_2 E_1^s) \cdot \delta D_1 -
(B_1 \Delta/E_1^s)  \cdot \delta E_2 +
(w \,sin[\psi_D - \Omega \tau]/E_1^s) \cdot \delta E_{2 \tau}
\end{eqnarray}
\begin{eqnarray}
\nonumber
\dot{\delta D_1}= (d_\| E_1^s/A )  \cdot \delta E2-  (d_\|^a + \frac{ (E_1^s)^2 d_\|}{\gamma_m^2}) \cdot \delta D1
\end{eqnarray}
\begin{eqnarray}
\nonumber
\dot{\delta \psi_s}=-B_3  \psi_1 + B_3 \cdot \delta \psi_2 - B_3 \cdot \delta \psi_s
\end{eqnarray}
\begin{eqnarray}
\nonumber
\dot{\delta D_0}=-(2  d_\| E_1^s/A) \cdot \delta E_1 - d_\| (1 + (E_1^s)^2/\gamma_m^2) \cdot \delta D_0,
\end{eqnarray}
\end{widetext}
where $B_1=w/\gamma_m$; \quad
$B_2=\frac{E_1^s}{(d_\|^a/d_\|+(E_1^s)^2) \gamma_m^2}$; \quad
$B_3=d_\|^a+ (E_1^s)^2 d_\| $; \quad
$B_4=A/\gamma_m^2$
These equations are a long-studied linear delayed equation \cite{Eltsgol}. We shall look for the solution having the form $\{ \delta E_1(t),\delta E_2(t),\delta \psi_2(t),\delta \psi_2(t),\delta D_1(t),\delta \psi_s(t),\delta D_0(t) \}=\overrightarrow{a} e^{p\, t}$, where $\overrightarrow{a}$ is a seven-dimensional constant vector, and $p$ is a constant. For that purpose it is necessary to solve the algebraic equation for the determinant, just as for common linear equations. For the delayed differential equations this algebraic characteristic equation is quasi-polynomial and has the form
\begin{equation}
\label{eq:quasi polinom}
P_0(p)+w^2(P_1(p)+P_2(p)\,e^{-\tau \,p}+P_3(p)\,e^{-2\,\tau\,p})=0
\end{equation}
 accurate to $w^2$. Here $P_{0,1,2,3}(p)$ are polynoms, $w^2<<1$ and $\tau>> d_\|$.
It is known that the steady point of nonlinear equation is stable if and only if the linearizated equation have only the decreasing solutions ($Re[p]<0$ for all $p$).
In the physically-interesting systems $d_\|<<1$, $A>1$, $\Delta^2<<1$. The case without external feedback \cite{Zeghlache} corresponds to $w=0$. Then (\ref{eq:quasi polinom}) has $Re[p]<0$ for $\Delta^2<\frac{d_\| \tilde A^2}{\tilde A-1}$.

It will be shown that the feedback force necessary for switching is decrease to zero when $\Delta^2 \rightarrow \frac{d_\| \tilde A^2}{\tilde A-1}$, so $\Delta^2$ and $d_\|$ have the same order of smallness. The values  $p^2$ and $d_\|$ have the same order of smallness too because $p_{w=0} \approx \sqrt{d_\|}(-i \sqrt{A-1}-\frac{A \sqrt{d_\|}}{2}+\Delta \frac{\sqrt{A-1}}{2})$ for  w=0.The series expansion in parameters $d_\|$ and $\Delta^2$ shows that $P_3(\imath p_i)>>P_2(\imath p_i)$.

Let`s find the constant $w=w_0$, when the solution of (\ref{eq:quasi polinom}) has $Re[p]=0$, $p=0+\imath p_i$.  We consider

\begin{equation}
\label{eq:quasi polinom_simple}
P_0(i p_i)+w^2(P_1(i p_i)+P_3(i p_i)\,e^{-i 2 \tau \,p_i})=0,
\end{equation}

where the real numbers $w^2$ and $p_i$ are unknown.

In the case $\tau>>1$ it is incorrect to use the standard perturbation theory with a small parameter $w^2$ to solve this algebraic equation, because of the fast oscillation of the small correction $w^2(P_1(i p_i)+P_3(i p_i)\,e^{-i 2 \tau \,p_i})$. That`s why we used special method
described in Appendix \ref{AppreshBOSC}.

\section{}
\label{AppreshBOSC}
To explain the solution method of the equation (\ref{eq:quasi polinom_simple}), let us consider the simple equation $sin(b x)=a x$ ,where $b>>a$, so $sin(b x)$ is a fast-oscillating function. This equation has no roots $x_0$, for  $a x_0>1$. If $b>>a$ the maximum root $x_{max}$ satisfies the equation $x_{max} a \approx 1$. The principal idea here is to substitute the fast oscillating function by its envelope.

As a bit more complicated example, we consider the equation
 \begin{equation}
\label{eq:prostoyPrimer}
p^2+b\,p+c+w\,((p^2+p)e^{-\tau p}+2)=0
\end{equation}
 for the complex p.  Here $b<<c$, $\tau>>1$, $b>0 ; c>0$.  The formulation of this problem is completely analogous to (\ref{eq:quasi polinom}).

 For $w=0$ this equation has two roots with a negative real part. Our purpose is to find  the value of $w_{crit}$, when the real part becomes zero. It means that p has the form $p=i\,q$, where q is a real number, so the equation can be rewritten as $f(q)=-q^2+i\,b\, q+c+w ((-q^2+i q)e^{-i\,\tau\,q}+2)$ and $f(q)=0$. Now we can consider the left part of the equation $f(q)$  as a two dimensional parametric curve:
 $$
x(q)=Re[f(q)]=-q^2+c+w(Re[ (-q^2+i q)e^{-i\,\tau\,q}]+2)
$$
$$
y(q)=Im[f(q)]=b q+w(Im[ (-q^2+i q)e^{-i\,\tau\,q}]).
$$
The problem has been reduced to the problem of the finding the value $w$ for which the parametric curve passes through the origin(x=0;y=0). In the case w=0 the smooth curve $f(q)$ is passing near the zero. When $w\neq0$ the typical curve is shown in the fig. \ref{fig:curve}. After the averaging this curve over the fast oscillating period one obtains:
 \begin{figure}[]
\center{\includegraphics[width=1\linewidth]{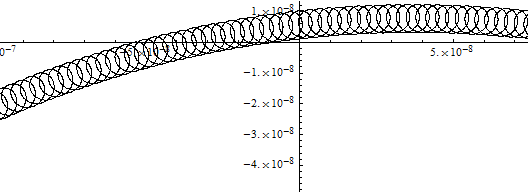}}
\caption{ }
\label{fig:curve}
\end{figure}
  $$
\bar{x}(q)=-q^2+c+2 w
$$
$$
\bar{y}(q)=b q.
$$
The fast oscillating part is:
$$
x_{osc}(q)=Re[w (-q^2+i q)e^{-i\,\tau\,q}]
$$
$$
y_{osc}(q)=Im[w (-q^2+i q)e^{-i\,\tau\,q}]
$$

So we can build two envelop curves and the whole curve is confined between them as it is shown in fig \ref{fig:curve} Our purpose is to find such a $w$ for which the coordinate origin will be between these envelopes. The distance between mean and envelope curves is a smooth function of q and could be found as $d=Abs[w (-q^2+i q)]$. In our case the point on the mean curve nearest to the origin, is the root of the equation $q_0^2-c\approx0$. Using this geometry considerations one obtains that $w_{crit}=\frac{b}{\sqrt{c+1}}$


\begin{thebibliography}{99}
%
\bibitem{ActSt1} L. Mashal, R. M. Nguimdo and Van der Sande, IEEE J. Quantum Elect.  \textbf{49}, 790  (2013)

     \bibitem{ActSt2}N. Li, W. Pan, S. Xiang et al. Opt. Laser Technol. \textbf{53}, 45  (2013)
     \bibitem{ActSt3}S.T. Kingni,P. Woafo, J. Mod. Optic \textbf{60}, 869 (2013)
     \bibitem{ActSt4}N. Li, W. Pan, S. Xiang et al., Appl. Pptics  \textbf{52} , 1523 (2013)
     \bibitem{ActSt5}R.M. Nguimdo,G. Verschaffelt,J. Danckaert et al., Opt. express  \textbf{20}, 28603 (2012)

  \bibitem{Zeghlache} H .Zeghlache, P. Mandel ,  Phys. Rev. A \textbf{37}, 470 (1988)

  \bibitem{Noise1} S. Surendra Singh and L. Mandel ,  Phys. Rev. A  \textbf{20}, 2459 (1979)

   \bibitem{Noise2} T. H. Chyba ,  Phys. Rev. A  \textbf{40}, 6327 (1989)

 \bibitem{Tang} D.Tang, R. Dykstra, Opt. Commun.  \textbf{126}, 318 (1996)

 \bibitem{Gas2}  Klische and C. O. Weiss,  Phys. Rev. A  \textbf{31}, 4049 (1985)

\bibitem{Dye} R. Roy and L. Mandel, Opt. Commun.  \textbf{34}, 133(1980)

\bibitem{Nd-YAG}S. Schwartz, G. Feugnet, E. Lariontsev, and Jean-Paul Pocholle, Phys. Rev. A  \textbf{76}, 023807 (2007)

\bibitem{conc}I.V. Ermakov, G. Van der Sande, J. Danckaert , Commun. Nonlinear Sci. Numer. Simul. \textbf{17}, 4767 (2012)

  \bibitem{sorel}  M. Sorel and P. J. R. Laybourn Opt. Lett.  \textbf{27}, 1992 (2002)

\bibitem{Haken}H. Haken \textit{Laser Light Dynamics} (Institute fur Theoretische Physik, Stuttgart, 1985), Vol. 2, p. 121


\bibitem{nota} P. Mandel, G.P. Agrawal, Opt. Commun.   \textbf{52}, 269

\bibitem{Eltsgol} L.E. Elsgolts, S.B. Norkin. \textit{Introduction to the theory and application of differential equations with deviating arguments}(Academic,New-York, 1973), p. 119

\bibitem{Diffus} I.V. Koryukin,P.A. Khandokhin,Y.I. Khanin,  Quantum Electron.   \textbf{25}, 1045 (1995)


\bibitem{quantDots}P. G. Eliseev, H. Li, A. Stintz, G. T. Liu, T. C. Newell Appl. Phys. Lett. \textbf{77}, 262 (2000)


\end{thebibliography}
\end{document}